\documentclass[twocolumn,prc,showpacs,showkeys]{revtex4}

\usepackage{amsfonts}
\usepackage{amsmath}
\usepackage{amssymb}
\usepackage{graphicx}
\usepackage{rotating}

\providecommand{\U}[1]{\protect\rule{.1in}{.1in}}
\providecommand{\U}[1]{\protect\rule{.1in}{.1in}}
\providecommand{\U}[1]{\protect\rule{.1in}{.1in}}

\begin{document}

\preprint{}
\title{On low-energy nuclear reactions}
\author{P\'{e}ter K\'{a}lm\'{a}n}
\author{Tam\'{a}s Keszthelyi}
\affiliation{Budapest University of Technology and Economics, Institute of Physics,
Budafoki \'{u}t 8. F., H-1521 Budapest, Hungary\ }
\keywords{nuclear reactions: specific reactions: general, quantum mechanics,
fusion reactions}
\pacs{24.90.+d, 03.65.-w, 25.60.Pj}

\begin{abstract}
Based on our recent theoretical findings (Phys. Rev. C \textbf{99}, 054620
(2019)) it is shown that proton and deuteron capture reactions of extremely
low energy may have accountable rate in the case of all elements of the
periodic table. Certain numerical results of rates of nuclear reactions of
two final fragments of extremely low energy are also given. New way of
thinking about low-energy nuclear reactions (LENR) phenomena is suggested.
Possible explanations for the contradictory observations announced between
1905-1927 and possible reasons for negative results of 'cold fusion'
experiments published recently by the Google-organized scientific group
(https://www.nature.com/articles/s41586-019-1256-6) are given.
\end{abstract}

\volumenumber{number}
\issuenumber{number}
\eid{identifier}
\date[Date text]{date}
\received[Received text]{date}
\revised[Revised text]{date}
\accepted[Accepted text]{date}
\published[Published text]{date}
\startpage{1}
\endpage{}
\maketitle

\section{Introduction}

Between 1905 and 1927, before the birth of quantum mechanics, such authors
as e.g. J. J. Thomson, W. Ramsay, J. N. Collie and H. Patterson debated the
possibility of creating $H$, $He$ and $Ne$ in gas discharges \cite%
{HeNe1905-1927}. The discussion was started by C. Skinner in \textit{The
Physical Review }\cite{Skinner}\textit{\ }and other articles appeared in
periodicals like \textit{Nature, Proceedings of the Royal Society of London A%
} and \textit{Science.} The case remained open until nuclear physics
answered it in the negative showing that the probability of reacting two
free nuclei of like electric charge at room temperature is below any
measurable value. However, the problem came to the fore again in a modified
form 30 years ago when it was reported \cite{FP1} that during electrolysis
excess heat and extra neutrons were generated in deuterized metal hydrides,
that could be attributed to and interpreted as evidence of nuclear fusion.
After a short period of heated debate scepticism has overcome the scientific
community \cite{Huizenga} related to the phenomenon called 'cold fusion' at
the time. Researchers were soon divided according to their opinion on the
possibility of 'cold fusion' and in spite of the prevailing negative
attitude of most (nuclear) physicists the quest for low-energy (temperature)
nuclear reactions (LENR) has been going on during the last three decades 
\cite{Storms3}-\cite{Storms1}. Unfortunately, experimental data are often
controversial and not clear enough to describe the essential features of
LENR. The situation requires theoretical guide lines to proposals of
experiments of new type. The main cause for the lack of proper guidelines
has been the missing theoretical answer to the basic question: how the
repelling nuclei can come close enough together to be able to take part in
nuclear processes if the kinetic energy $\varepsilon $\ of relative motion
is very small. (The problem is usually connected to the problem of tunneling
through a large potential barrier caused by Coulomb repulsion.)

Recently however, the necessary step to solve the basic problem was taken 
\cite{KK}. In what follows the theoretical results of \cite{KK}\ are
summarized.

From the point of view of LENR the magnitude of the wavefunction of reacting
particles in nuclear range is crucial. The original Coulomb solution \cite%
{Alder} yields with decreasing $\varepsilon $ vanishing contact probability
density in the nuclear volume that in this case leads to disappearing rate
of nuclear reactions. In \cite{KK} the problem of forbidden nuclear
reactions was compared to the forbidden optical transitions, particularly to
the case of the forbidden $2s_{1/2}-1s_{1/2}$ transition of the hydrogen
atom where the transition of the electron is accompanied by the emission of
two photons that can be traced by second order perturbation calculation \cite%
{Bethe}. Accordingly in \cite{KK} it is pointed out that '\textit{any
perturbation} can mix states with small but finite amplitude to the initial
state resulting in finite cross section (and rate)' also in the case of
originally vanishing contact probability. The statement was illustrated by
modification of nuclear reactions due to impurities in a gas mix of atomic
state. Standard time independent perturbation calculation of quantum
mechanics \cite{Landau} was used to determine the change of the wavefunction
of reacting particles in nuclear range due to their Coulomb interaction with
impurity leading to a nonvanishing wavefunction and contact probability
density in nuclear volume even if their $\varepsilon $ goes to zero. The
rate of nuclear processes discussed is proportional to $n_{1}n_{2}n_{3}$,
where $n_{j}$ is the number density of the participating particles $j$. It
was also obtained in \cite{KK} that this proportionality remains valid at
the surface of solids where in this case $n_{1}=N_{c}/v_{c}$ with $v_{c}$
the volume of unit cell and $N_{c}$ the number of atoms in it. Consequently,
the higher the product $n_{1}n_{2}n_{3}$ is the better the process works.
The results in \cite{KK} are general and the model is applicable in the
cases of gas discharges, heated metal-gas systems and electrolysises where
LENR phenomena usually happen.

The cross sections of reactions of%
\begin{equation}
_{z_{1}}^{A_{1}}V+\text{ }_{z_{2}}^{A_{2}}w+\text{ }_{z_{3}}^{A_{3}}X%
\rightarrow \text{ }_{z_{1}}^{A_{1}}V^{\prime }+\text{ }_{z_{4}}^{A_{4}}Y+%
\Delta  \label{Reaction 3}
\end{equation}%
and 
\begin{equation}
_{z_{1}}^{A_{1}}V+\text{ }_{z_{2}}^{A_{2}}w+\text{ }_{z_{3}}^{A_{3}}X%
\rightarrow \text{ }_{z_{1}}^{A_{1}}V^{\prime }+\text{ }_{z_{4}}^{A_{4}}Y+%
\text{ }_{z_{5}}^{A_{5}}W+\Delta  \label{Reaction 4}
\end{equation}%
were determined. Here $z_{j}$ and $A_{j}$ are charge and mass numbers and $%
\Delta $ is the reaction energy which is the difference between the sum of
the initial and final mass excesses \cite{Shir}. Numerical evaluations
indicate that the rate of reactions that were expected to be negligible
small, may become of accountable magnitude due to Coulomb assistance by the
impurity $_{z_{1}}^{A_{1}}V$.

Based on the theoretical findings presented in \cite{KK} and the further
results to be disclosed here, in this paper we are going to propose a new
approach to understanding the divers experimental results in the field of
LENR. Especially, the explanation of negative results concerning the
existence of 'cold fusion' published and disclosed recently by the
Google-group \cite{Nature}, needs placing 'cold fusion' into a broader
context of LENR. We also try to reconcile the occasional transmutations
observed in early experiments \cite{HeNe1905-1927} and point out the
possible reasons for the failures of recent 'cold fusion' experiments \cite%
{Nature}. It will be discussed in more detail in the second part of the
paper.

\section{Cross section calculation}

The cross section $\sigma _{23}^{\left( 2\right) }$ of processes $\left( \ref%
{Reaction 3}\right) $ and $\left( \ref{Reaction 4}\right) $ can be
determined as

\begin{equation}
v_{23}\sigma _{23}^{\left( 2\right) }=n_{1}S_{\text{reaction}},
\label{v23sigma23-2}
\end{equation}%
where $v_{23}$ is the relative velocity of particles $2$ and $3$. In the
case of reaction $\left( \ref{Reaction 4}\right) $ $S_{\text{reaction}}$ can
be deducted from astrophysical factors $S\left( \varepsilon \right) $ from
Eqs.(27) and (28) of \cite{KK} with the aid of $S\left( 0\right) $ in the
long wave approximation. Since the calculated $S_{\text{reaction}}(\sim
z_{1}^{2}S\left( 0\right) )$ values of reactions of two final fragments are
connected to experimental observations they are nuclear model independent.
The rate and power densities are $r_{\text{reaction}}=n_{1}n_{2}n_{3}S_{%
\text{reaction}}$ and $p_{\text{reaction}}=r_{\text{reaction}}\Delta
=n_{1}n_{2}n_{3}S_{\text{reaction}}\Delta $, respectively. (In the case of
reactions $^{15}N(p,\alpha _{0,1})^{12}C$ the $S(\varepsilon )$ function
determined by \cite{LaCognata} is applied with the aid of Eqs.(49)-(50) of 
\cite{KK2}. Here and below the subscripts 0 and 1 refer to ground and first
excited final nuclear states.)

Conversely, in the case of reaction $\left( \ref{Reaction 3}\right) $, which
is an assisted capture reaction, the calculations can not be based on the
astrophysical factors of reactions $_{z_{2}}^{A_{2}}w+$ $_{z_{3}}^{A_{3}}X%
\rightarrow $ $_{z_{4}}^{A_{4}}Y+\gamma $ since it is governed by
electromagnetic interaction while the nuclear part of $\left( \ref{Reaction
3}\right) \emph{\ }$is governed by strong interaction and therefore the
cross section calculations of $\left( \ref{Reaction 3}\right) $\ are
strongly nuclear model dependent. In \cite{KK} the cross section of $%
_{z_{1}}^{A_{1}}V+$ $p+$ $d\rightarrow $ $_{z_{1}}^{A_{1}}V^{\prime }+$ $%
^{3}He+5.493$ MeV was determined in a very simple model with an assisting
nucleus of mass and charge numbers $A_{1}$ and $z_{1}$. The result indicates
that the most significant increase of the cross section caused by Coulomb
interaction of reacting particles $\left( 2\text{ and }3\right) $ with $%
_{z_{1}}^{A_{1}}V$ originates from factors $f_{23}$ which come from the
Coulomb solutions of particles $2$, $3$ of high $\varepsilon $. According to
perturbation calculation \cite{KK} these Coulomb states are mixed to the
state of $\varepsilon =0$ and factors $f_{23}$, which come from these
Coulomb solutions, are defined as%
\begin{equation}
f_{23}^{2}(s,\Delta )=\frac{2\pi \eta _{23}(s,\Delta )}{\exp \left[ 2\pi
\eta _{23}\left( s,\Delta \right) \right] -1}  \label{f23}
\end{equation}%
with%
\begin{equation}
\eta _{23}\left( s,\Delta \right) =z_{2}z_{3}\alpha _{f}\frac{a_{23}}{a(s)}%
\sqrt{\frac{m_{0}c^{2}}{2a_{14}\Delta }},  \label{etajk}
\end{equation}
where $a(s)=\left\vert -A_{3}\delta _{s,2}+A_{2}\delta _{s,3}\right\vert
/\left( A_{2}+A_{3}\right) $ \ with $s=2,3$. $m_{0}c^{2}=931.494$ MeV is the
atomic energy unit, $c$ is the velocity of light in vacuum, $\alpha _{f}$ is
the fine structure constant and $a_{jk}=A_{j}A_{k}/\left( A_{j}+A_{k}\right) 
$ is the reduced mass number of particles $j$ and $k$ of mass numbers $A_{j}$
and $A_{k}$ with rest masses $m_{j}=A_{j}m_{0}$, $m_{k}=A_{k}m_{0}$. The
cross section and the rate are proportional to a sum which contains terms
proportional to products $f_{23}(s,\Delta )f_{23}(s^{\prime },\Delta )$.

In the case of proton capture ($z_{2}=1$, $A_{2}=1$) the $s=2$ case with $%
a(2)=A_{3}/\left( 1+A_{3}\right) $ gives the largest $f_{23}$ value
therefore only the leading $f_{23}^{2}(2,\Delta )$ will be studied on. (One
should remember that $f_{23}^{2}(2,\Delta )$ must be compared to $\exp \left[
-2\pi \eta _{23}\left( \varepsilon \right) \right] $ which comes from cross
section $\sigma \left( \varepsilon \right) =S\left( \varepsilon \right) \exp %
\left[ -2\pi \eta _{23}\left( \varepsilon \right) \right] /\varepsilon $ of
usual nuclear reactions between charged particles $2$ and $3$ of charge
numbers $z_{2}$ and $z_{3}$ \cite{Angulo} where $S\left( \varepsilon \right) 
$ is the astrophysical $S$-factor, $\eta _{23}\left( \varepsilon \right)
=z_{2}z_{3}\alpha _{f}\left[ a_{23}m_{0}c^{2}/\left( 2\varepsilon \right) %
\right] ^{1/2}$ is the Sommerfeld parameter and $\varepsilon $ is the
kinetic energy taken in the center of mass coordinate system. If $%
\varepsilon \rightarrow 0$ then $\sigma \left( \varepsilon \right) $ and the
rate disappear.)

\section{Numerical results}

\begin{table}[tbp]
\tabskip=8pt 
\centerline {\vbox{\halign{\strut $#$\hfil&\hfil$#$\hfil&\hfil$#$
\hfil&\hfil$#$\hfil&\hfil$#$\hfil&\hfil$#$\cr
\noalign{\hrule\vskip2pt\hrule\vskip2pt}
Reaction&S_{reaction}&\Delta&r_{reaction}\cr
\noalign{\vskip2pt\hrule\vskip2pt}
S(\varepsilon) & & & \cr
\noalign{\vskip2pt\hrule\vskip2pt}
^{15}N(p,\alpha_0)^{12}C & 8.20\times10^{-51} & 4.965 &1.61\times10^{11} \cr
^{15}N(p,\alpha_1)^{12}C & 1.63\times10^{-51} & 0.526 &3.19\times10^{10} \cr
\noalign{\vskip2pt\hrule\vskip2pt}
S(0)=100 & & & \cr
\noalign{\vskip2pt\hrule\vskip2pt}
^{23}Na(p,\alpha_0)^{20}Ne& 5.59\times10^{-52} & 2.377 &1.10\times10^{10} \cr
^{23}Na(p,\alpha_1)^{20}Ne & 4.43\times10^{-52} & 0.743 & 8.68\times10^{9} \cr
\noalign{\vskip2pt\hrule\vskip2pt}
S(0)=3.5\times10^{7}& & & \cr
\noalign{\vskip2pt\hrule\vskip2pt}
^{14}N(^{7}Li,\alpha_0)^{17}O & 9.71\times10^{-52} & 16.155& 1.90\times10^{10} \cr
^{14}N(^{6}Li,\alpha_0)^{16}O & 4.92\times10^{-51} & 19.262& 9.64\times10^{10} \cr
^{16}O(^{7}Li,\alpha_0)^{19}F & 2.43\times10^{-53} & 9.233& 4.76\times10^{8} \cr
^{16}O(^{6}Li,\alpha_0)^{18}F & 2.13\times10^{-53} & 6.051& 4.18\times10^{8} \cr
^{17}O(\alpha_0,n)^{20}Ne & 3.46\times10^{-54} & 0.587& 6.79\times10^{7} \cr
\noalign{\vskip2pt\hrule\vskip2pt}
S(0)=1.0\times10^{16}& & & \cr
\noalign{\vskip2pt\hrule\vskip2pt}
^{12}C(^{12}C,\alpha_0)^{20}Ne & 2.24\times10^{-58} & 4.617 &4390 \cr
^{12}C(^{12}C,\alpha_1)^{20}Ne & 4.43\times10^{-63} & 2.983 &0.087 \cr
^{12}C(^{12}C,p_0)^{23}Na & 8.11\times10^{-67} & 2.241 &1.60\times10^{-5} \cr
^{12}C(^{12}C,p_1)^{23}Na & 4.39\times10^{-70} & 1.801 &8.61\times10^{-9} \cr
\noalign{\vskip2pt\hrule\vskip2pt\hrule}}}}
\caption{$S_{\text{reaction}}$(in cm$^{6}$s$^{-1}$) and rate density $r_{%
\text{reaction}}$(in cm$^{-3}s^{-1}$) of $Ni$ (of mass number $58$) assisted
reactions with two final fragments. $\Delta $(in MeV) is the energy of the
reaction. $S(0)$(in MeVb) is the astrophysical factor at $\protect%
\varepsilon =0$, where $\protect\varepsilon $ is the center of mass kinetic
energy. $r_{\text{reaction}}=n_{1}n_{2}n_{3}S_{\text{reaction}}$ is the rate
density that is calculated with $n_{1}n_{2}n_{3}=1.861\times 10^{61}$ cm$%
^{-9}$. The subscripts 0 and 1 refer to ground and first excited final
nuclear states. The energies (spin$^{parities}$) of the considered excited
states are: $^{12}C$, $4.439$ MeV $(2^{+})$; $^{20}Ne$, $1.634$ MeV $(2^{+})$
and $^{23}Na$, $0.4400$ MeV$(5/2^{+})$. $S(\protect\varepsilon )$ is taken
from \protect\cite{LaCognata}, $S(0)=3.5\times 10^{7}$ MeVb, $S(0)=100$ MeVb
are taken from \protect\cite{Angulo}, \protect\cite{Fisher}, respectively.
Furthermore, the $S(0)=1.0\times 10^{16}$ MeVb \protect\cite{Spillane} value
seems to be an under-estimation of $S(0)$ \protect\cite{Tumino}.}
\end{table}

\begin{table}[tbp]
\tabskip=8pt 
\centerline {\vbox{\halign{\strut $#$\hfil&\hfil$#$\hfil&\hfil$#$
\hfil&\hfil$#$\hfil&\hfil$#$\hfil&\hfil$#$\cr
\noalign{\hrule\vskip2pt\hrule\vskip2pt}
Reaction&S_{reaction}&\Delta&r_{reaction}\cr
\noalign{\vskip2pt\hrule\vskip2pt}
S(0)=100 \cr
\noalign{\vskip2pt\hrule\vskip2pt}
^{16}O(d,\alpha_0)^{14}N & 1.86\times10^{-51} & 3.011 &3.66\times10^{10} \cr
^{17}O(d,\alpha_0)^{15}N &2.11\times10^{-51} & 9.800 &4.15\times10^{10} \cr
^{18}O(d,\alpha_0)^{16}N & 1.96\times10^{-51} & 4.245 &3.85\times10^{10} \cr
\noalign{\vskip2pt\hrule\vskip2pt}
S(0)=3.5\times10^{7} \cr
\noalign{\vskip2pt\hrule\vskip2pt}
^{16}O(d,\alpha_0)^{14}N & 6.52\times10^{-46} & 3.011 &1.28\times10^{16} \cr
^{17}O(d,\alpha_0)^{15}N & 7.40\times10^{-46} & 9.800 &1.45\times10^{16} \cr
^{18}O(d,\alpha_0)^{16}N & 6.86\times10^{-46} & 4.245 &1.35\times10^{16} \cr
\noalign{\vskip2pt\hrule\vskip2pt\hrule}}}}
\caption{$S_{\text{reaction}}$(in cm$^{6}$s$^{-1}$) and rate density $r_{%
\text{reaction}}$(in cm$^{-3}s^{-1}$) of $Pt$ (of mass number $195$)
assisted reactions with two final fragments. $\Delta $(in MeV) is the energy
of the reaction. $r_{\text{reaction}}=n_{1}n_{2}n_{3}S_{\text{reaction}}$ is
the rate density that is calculated with $n_{1}n_{2}n_{3}=1.861\times
10^{61} $ cm$^{-9}$. $S(0)$(in MeVb) is the astrophysical factor at $\protect%
\varepsilon =0$, where $\protect\varepsilon $ is the center of mass kinetic
energy. $S(0)=3.5\times 10^{7}$ MeVb, $S(0)=100$ MeVb are taken from 
\protect\cite{Angulo}, \protect\cite{Fisher}, respectively.}
\end{table}

Capture reactions of type $\left( \ref{Reaction 3}\right) $ will have cross
section $\sigma _{23}^{\left( 2\right) }$ of accountable magnitude if $\eta
_{23}<1$ (if $\eta _{23}=1$ then $f_{23}^{2}=0.012$). In the case of proton
capture using $\eta _{23}(2,\Delta )$ the $\eta _{23}<1$\ leads to condition 
$z_{3}\lesssim \left[ 2a_{14}\Delta /\left( \alpha _{f}^{2}m_{0}c^{2}\right) %
\right] ^{1/2}$. In the case of heavy $1$, $3$ and $4$ particles, $%
A_{1}\simeq A_{3}\simeq A_{4}$ and $a_{14}\simeq A_{4}/2\gtrsim z_{4}\simeq
z_{3}$ one has $z_{3}\lesssim 2\Delta /\left( \alpha
_{f}^{2}m_{0}c^{2}\right) =40.30\times \Delta ($in MeV$)$. For $d$-capture
the condition modifies as $z_{3}\lesssim \Delta /\left( 2\alpha
_{f}^{2}m_{0}c^{2}\right) =10.08\times \Delta ($in MeV$)$. All this means
that e.g. even the $^{238}U+$ $d+$ $^{235}U\rightarrow $ $^{238}U^{\prime }+$
$^{237}Np+9.182$ MeV $d$-capture reaction can happen with a non-negligible
probability (rate). As a result we are faced with plenty of possible $p$-
and $d$-capture reactions of type $\left( \ref{Reaction 3}\right) $.
Obviously the $_{z_{1}}^{A_{1}}V+$ $d+$ $d\rightarrow $ $_{z_{1}}^{A_{1}}V^{%
\prime }+$ $^{4}He+23.845$ MeV reaction may also happen. An essential
consequence of the possibility of $p$- and $d$-capture reactions of type $%
\left( \ref{Reaction 3}\right) $ is the possibility of nuclear transmutation
of all the elements of the periodic table.

The results of our $S_{\text{reaction}}$ and rate density calculations can
be found in Tables I and II. The energies (spin$^{parities}$) of the
considered excited states are: $^{12}C$, $4.439$ MeV $(2^{+})$; $^{20}Ne$, $%
1.634$ MeV $(2^{+})$ and $^{23}Na$, $0.4400$ MeV $(5/2^{+})$. $S(\varepsilon
)$ is taken from \cite{LaCognata}, $S(0)=3.5\times 10^{7}$ MeVb, $S(0)=100$
MeVb are taken from \cite{Angulo}, \cite{Fisher}, respectively. Furthermore,
the $S(0)=1.0\times 10^{16}$ MeVb \cite{Spillane} value seems to be an
under-estimation of $S(0)$ \cite{Tumino}.

It was found that not only in the case of the earlier investigated reactions
(the reactions from $d(d,n)^{3}He$ up to $^{11}B(p,\alpha )^{8}Be$, see
Table I of \cite{KK}) but in view of the results obtained above for
reactions discussed here (reactions from $^{15}N(p,\alpha _{0})^{12}C$ up to 
$^{12}C(^{12}C,\alpha _{1})^{20}Ne$, see Table I here and the reactions of
Table II) the reaction rate can have accountable magnitude if the number
densities $n_{j}$ $\left( j=1,2,3\right) $ of the initial particles reach
appropriately high values.

\section{Interpretation of LENR experiments of contradictory and negative
results}

\subsection{Experiments of contradictory results}

Now, as a first example we discuss the early contradictory observations
related to $H$, $He$ and $Ne$ creation in gas discharges \cite{HeNe1905-1927}%
. In the experiment by Skinner \cite{Skinner} a testing electrode from $NaK$
alloy remained in open connection with the discharge tube where the test
(metal) electrodes operated, therefore $^{23}Na-$$^{4}He$ interaction could
happen. Due to this interaction assisted by $Na$ atoms of the alloy the $%
^{23}Na+$ $^{23}Na+$ $^{4}He\rightarrow $ $^{23}Na^{\prime }+p+$ $%
^{26}Mg+1.821$ MeV reaction could take place. Calculating the rate taking $%
n_{1}=n_{3}=1\times 10^{23}$ cm$^{-3}$ for $Na$ atom number density in the
alloy, $n_{2}=1.02\times 10^{17}$ cm$^{-3}$ which corresponds to the applied 
$^{4}He$ content (pressure) and using $S(0)=3.5\times 10^{7}$ MeVb,$\ S_{%
\text{reaction}}=5.41\times 10^{-55}$ cm$^{6}$s$^{-1}$ and $r_{\text{reaction%
}}=2.08\times 10^{8}$ cm$^{-3}$s$^{-1}$ rate density are obtained. The
disappearance of $H$ at the $Al$ anode can be put down to the $Al$ assisted $%
^{27}Al+$ $^{27}Al+p\rightarrow $ $^{27}Al^{\prime }$ $+$ $^{4}He+$ $%
^{24}Mg+1.600$ MeV reaction. Moreover, the $^{23}Na+$ $^{23}Na+$ $%
p\rightarrow $ $^{23}Na^{\prime }+$ $^{4}He+$ $^{20}Ne$ reactions (see for
similar reactions Table I) could also work if the glass from which the
discharge tube was made had some $Na$ content.

Now we consider the occasional $He$ and $Ne$ creation in gas discharges \cite%
{HeNe1905-1927}. Discharge tubes are constructed sometimes from quartz and
sometimes from borosilicate that may have some $Na$ content too. During
discharge protons bomb the wall of the experimental apparatus due to
ambipolar diffusion, and if the discharge tube was made from borosilicate
glass with $Na$ content then reactions $^{10}B(p,\ \alpha _{0})^{7}Be$,$%
^{11}B(p,\ \alpha _{0})^{8}Be$ (the $^{8}Be$ spontaneously decays to $%
2\alpha $) and $^{23}Na(p,\alpha _{0,1})^{20}Ne$ could happen producing $He$
and $Ne$. A fact that can explain the mystery of their occasional production
since in the absence of $B$ and $Na$, which is the case when using quartz,
the nuclear reactions do not happen. In addition, if the discharge tube has
constriction as was the case of the apparatus of \cite{Collie} then the
effect caused by ambipolar diffusion is stronger.

\subsection{Experiments of negative results of Google-organized research
group}

In \cite{Nature}, three principal directions of research were specified:
highly hydrated metals, calorimetry under extreme conditions and low-energy
nuclear reactions. (This later terminology was to define a special pulsed
deuterium plasma device \cite{Schenkel} which was applied to induce nuclear
reactions of low energy.) However, nuclear transmutation \cite{Storms3}-\cite%
{Storms1}, which is the most important phenomenon connected to LENR, was
missing. It was shown above that nuclear transmutation is possible for all
the elements of the periodic table therefore it is expected that traces of
it must be present in all LENR observations. There exist very sensitive
methods which are capable to determine small amounts of changes of chemical
composition of materials and show the appearance of nuclear transmutation.
Thus omission of search for nuclear transmutation is the main fault in the
program of \cite{Nature}.

\subsubsection{Highly hydrated metals}

The $^{4}He$ production in electrolysis seemed to be accompanied by extra
heat production, that was attributed to $d+$ $d\rightarrow $ $^{4}He+\gamma $
reaction with the implicit expectation of occurrence of $d(d,t)p$ and $%
d(d,n)^{3}He$ reactions bringing up the problem of the missing $\gamma $, $n$
and $t$. However, as can be seen from Table I here and in \cite{KK} and
Table II $^{4}He$ production can occur in many other reactions too. The sum
of their energy production may significantly exceed the sole energy
production of $_{z_{1}}^{A_{1}}V+d+$ $d\rightarrow $ $^{4}He+$ $%
_{z_{1}}^{A_{1}}V^{\prime }$ (the appearance of which in itself can partly
solve the problem of missing $\gamma $ in $^{4}He$ production) and the rate
of $n$, $t$ and $^{3}He$ productions in assisted $d(d,t)p$ and $d(d,n)^{3}He$
reactions may be considerably smaller than the total rate of $^{4}He$
production. In electrolysis-type experiments one usually focuses on the
cathode and searching for nuclear reactions attached to the anode is
omitted. In order to demonstrate the possible importance of nuclear
reactions which may be connected to the anode, the rates of a few $\left(
d,\alpha \right) $ reactions of type $\left( \ref{Reaction 4}\right) $
assisted by $Pt$($A_{1}=195$, $z_{1}=78$) at two possible $S(0)$ values are
listed in Table II. Comparing the obtained rate numbers with the rate
numbers of Table I one can recognize that the rates of Table II are
approximately equal to or larger than the numbers of Table I which fact
calls attention to the possible importance of nuclear processes happening
near the surface of the anode. It is stated that large $D/Pd$ ratio of
loading $D$ in $Pd$ is necessary to observe effects \cite{Storms2}, \cite%
{McKubre}. Although this observation suggests that the most important
nuclear reactions take place in the loaded $Pd$, since other parameters
(e.g. current density of electrolysis) must also have extreme values in
order to reach the desired high $D/Pd$ ratio, more advantageous
circumstances may come about in other places of the experimental equipment
that permit other concurrent nuclear reactions of high power density to
start. [It is worth mentioning that e.g. the $_{z_{1}}^{A_{1}}V+$ $^{16}O+$ $%
p\rightarrow $ $^{13}N+$ $^{4}He+$ $_{z_{1}}^{A_{1}}V^{\prime }$ reaction is
endothermal ($\Delta =-5.218$ MeV) and the reaction energy ($\Delta =0.600$
MeV) of $_{z_{1}}^{A_{1}}V+$ $^{16}O+$ $p\rightarrow $ $^{17}F+$ $%
_{z_{1}}^{A_{1}}V^{\prime }$ capture reaction is small. These reactions can
take place in electrolysises in normal water.] In view of the above it must
be emphasized that any seemingly tiny detail of geometry and chemical
composition of structural elements can be crucial. Therefore precise and
very detailed documentation of the applied experimental apparatus is
required.

\subsubsection{Calorimetry in extreme conditions}

The calorimetry performed in extreme conditions in \cite{Nature} has been
focused on powdered $Ni-H-LiAlH_{4}$ systems. In these and similar
experiments metals ($Ni$ and $Pd$) were used in order to break the (mainly
two-atomic) molecules into atomic state. Since the mechanism works on the
surface of metals \cite{Kroes} it is tacitly assumed that the nuclear
reactions too take place very close to\ the metal surface. In the
experiments of the early 1990s \cite{Focardi}, \cite{Focardi 2}\ where $Ni$
rod $-$ $H$ gas systems were investigated ceramic holders were applied to
keep metal rods inside the experimental chamber. The chemical composition of
the ceramic holder is unknown but it is reasonable to assume that it may
have contained elements which took part in nuclear reactions and produced
measurable excess heat. In the experiments on powdered $Ni-H-LiAlH_{4}$
systems after 2011 $N_{2}$ gas buffer was also applied. (There was an $N_{2}$
gas container visible in an unpublished internet-picture so it is of great
probability that in order to avoid $H_{2}$ explosion buffer gas $N_{2}$ was
used.) The $^{15}N(p,\alpha _{0})^{12}C$, $^{15}N(p,\alpha _{1})^{12}C$, $%
^{14}N(^{7}Li,\alpha _{0})^{17}O$ and $^{14}N(^{6}Li,\alpha _{0})^{16}O$
reactions may contribute to large fraction of power generation which may
explain negative experimental results \cite{Nature} obtained without $N_{2}$
gas buffer in the active volume.

\subsubsection{Reactions induced by pulsed deuterium plasma beam}

Finally, a special device \cite{Nature}, \cite{Schenkel} was used to
investigate nuclear $dd$ reactions induced by pulsed deuterium plasma beam
bombarding palladium targets with deuterons producing more flux than that of
ion beams of commonly used low energy accelerators. However, since the
vacuum chamber contained deuterium gas ($D_{2}$) at about 1 torr the product 
$n_{1}n_{2}n_{3}$ remained much below the necessary $10^{61}$ cm$^{-9}$
order of magnitude value to reach accountable rate.

\section{Conclusion}

Based on the recent theoretical results \cite{KK} it was shown that a huge
number of nuclear reactions may have significant rate even if the kinetic
energy of the colliding particles is down to that of room temperature. The
necessary condition to reach accountable rate is that the product $%
n_{1}n_{2}n_{3}$, i.e. the product of the number density of the assisting
particles and the number densities of the reacting particles, should reach
appropriately high value. Since the participants of possible nuclear
reactions can come from the whole periodic table of elements, it is advised
to drop old stereotypes in thinking about LENR and a new approach is
necessary to understand what is going on in this field.

\end{document}